\documentstyle[preprint,aps]{revtex}
\begin{document}
\draft



\title{The Cu NQR Study of the Stripe Phase Local 
Structure in the Lanthanum Cuprates}

\author{G.B.Teitel'baum,$^1$ B.B\"uchner,$^2$ 
H.de Gronckel$^3$}

\address{$^1$Institute for Technical Physics, 
420029 Kazan, Russia\\
$^2$II Physikalisches Institut, Universit\"at 
zu K\"oln, D-50937 K\"oln,
Germany\\ $^3$IFF Forschungszentrum J\"ulich, 
D-52425 J\"ulich, Germany}

\date{\today }
\maketitle

\begin{abstract}

Using Cu NQR in Eu-doped $\rm La_{2-x}Sr_xCuO_4$\ 
we find 
the evidence of the pinned stripe phase at 1.3K 
for $0.08\leq x\leq 0.18$. The pinned fraction 
increases by one order of 
magnitude 
near hole doping $x=1/8$. The NQR lineshape reveals
three inequivalent Cu positions: i) sites in the 
charged stripe; 
ii) nonmagnetic sites outside the stripes; iii) 
sites with a 
magnetic moment of 0.29$\mu_B$ in the AF correlated 
regions.  
A dramatic change of the NQR signal for  $x > 0.18$ 
correlating
with the onset of bulk superconductivity corresponds to 
the depinning of the stripe phase. 

\end{abstract}


The recognition is growing that doping of the 
antiferromagnetic
(AF) insulating phase of a high-$T_c$ superconductor 
by holes has an 
explicit
topological character.  In fact, according to the 
time reversal symmetry the
segregation of charges to periodical domain walls 
(stripes) requires
 an
antiphase arrangement of the created AF domains \cite{3,2,3b}.
The first evidence for such stripe phase has been provided by 
neutron studies of
the low temperature tetragonal (LTT) phase of Nd-doped
$\rm La_{2-x}Sr_xCuO_4$\ \cite{1}.  A number of recent papers 
confirmed the
presence of stripe correlations in the other cuprates as 
well\cite{5,5a,5b}.
But in spite of the hot interest to the problem suprisingly 
little is known
about the local properties of the stripe structure.

In this Letter we report the results of the direct study of 
the stripe phase 
local structure
by  means of Cu NQR. The application of NMR and NQR for study 
of stripes
meets serious difficulties due to the slowing of the charge 
fluctuations down
to MHz frequency range which wipes out a large part of the 
nuclei from the
resonance.  An important breakthrough\cite{7} based on the 
quantitative
analysis of the fraction of the nuclei wiped out from the NQR 
brought insight
in the behaviour of the stripe phase order parameter both in 
cuprates \cite
{7} and nickelates \cite{7a}.

Unfortunately, based on wipeout effects alone, it is impossible 
to determine the
local structure of the stripes, i.e. the charge and internal 
magnetic field
distribution, and the values of the typical local parameters. 
 Such
information can be obtained only from the NQR analysis of the 
stripe phase
itself, which is possible after reappearance of the signal 
in the slow
fluctuation limit.  The pinning of stripes at low temperatures
 enables us to
take the advantages of the extreme sensitivity of Cu NQR to 
the local charge
and magnetic field distribution. In addition to the 
measurements at
temperatures of 1.3 K, this program could be realized easier 
for the LTT
structure, which is helpful for pinning of the stripe 
structure. This
structure was induced by doping with non-magnetic Eu rare-earth 
ions instead
of magnetic Nd ones (the ordering of Nd moments causes fast 
Cu nuclear
relaxation hindering the observation of Cu NQR). We expect 
that in the stripe
structure the different Cu sites will be inequivalent with 
respect to the NQR,
providing information on the local properties at given 
points of the
structure.

For our experiments we have chosen fine powders of 
the series
$\rm La_{2-x-y}Eu_ySr_xCuO_4$\ with variable
Sr content $x$ and fixed  Eu content $y =0.17$. The 
preparation of single
phase samples was described in \cite{8}. It was 
found \cite{8} that for
such Eu content the LTT phase is realized  
for $x>0.07$. For Sr concentrations
$x>0.12$ the ac-susceptibility and microwave absorption 
measurements reveale
the presence of superconductivity with  
$T_c = 6;\ 9;\ 14;\ 19;\ 18;\ 16;\
13$K for resp. $x= 0.12;\ 0.13;\ 0.15;\ 0.18;\ 0.20;\;0.22; \ 0.24$.
The  superconducting fraction is small for  
$x\leq 0.18$ and starting from
$x>0.18$ a transition to bulk superconductivity 
take place.
The NQR measurements are performed with the standard 
spectrometer in the
range 20 - 100 MHz. By lowering the temperature down 
to 1.3K we
succeed to observe the Cu-NQR spectra at all Sr 
concentrations.  Regarding
their NQR properties the samples should be separated 
into two groups.

The first one corresponds to Sr concentrations 
$x\leq 0.18$. The
superconducting fraction of these samples, if any, 
was rather small.  Each of
the spectra, which are very similar for $0.08\leq x\leq 0.18$, 
consists of
a broad line in the region from  20 MHz up to 80 MHz with 
an unresolved
peak between 30 and 40 MHz  (examples of some spectra 
are shown in Fig. 1).
The main distinctions of the spectra for different $x$  
are i) the integral intensity of the spectra, which is
peaked near $x=0.12$ (Fig. 2a), and ii) the  
temperatures below which it is
possible to observe them (for $x= 0.12; 0.13$  they are
observable even for the temperatures higher than 4.2K).

The second group of samples with $x>0.18$ showing 
bulk superconductivity
posesses completely different and much narrower 
NQR spectra (Inset to Fig.1),
which can also be observed at much higher temperatures.  
The intensity grows
with increasing $x$ from 0.18 (Fig. 2b).

Beginning the discussion with the  
$x\leq 0.18$ group, we first
consider the above mentioned complicated peak 
in the lineshapes. Since the
natural abundance ratio of  $^{63}$Cu and $^{65}$Cu 
is 2.235 and the ratio of
their quadrupole moments is 1.081 it is clear that 
this peak contains more
than one pair of  $^{63}$Cu and $^{65}$Cu signals.  
The possibility that such
a picture arises from one site due to the splitting 
of the signal by the
hyperfine field can be ruled out by the different behaviour of the relative
intensities of both components upon variation of $x$ 
and by their different echo 
decay times.  The gaussian fit to these  peaks reveals 
the existence of two independent  copper sites 
$\it{1}$ and $\it{2}$ (we use this notation in order 
to distinguish them from
the sites A and B known for the superconductors in 
the low temperature orthogonal
(LTO) phase \cite{Yasuoka}), having different NQR frequencies 
(Fig. 3).

To make the site assignment, we note that the NQR frequency 
is sensitive to the
the local hole concentration changing between 0.5 and 0 
hole per Cu atom
\cite{1}. In a linear approximation  we obtain that for 
the given 
$x$ the resonant frequency $\nu_Q$\ is connected with 
the local hole density $n(r)$\ via the relation
$\nu_Q(x,n)$ = $\nu_Q^0$ $-$  $\alpha$$x$ + $\beta$$n$  
with the empirical constants
$\alpha$ and $\beta$. 
The first term here is the NQR frequency for the 
compound with zero Sr
content, the second one is due to the negative shift 
caused by the 
contraction of Cu-O bond
length induced by the internal pressure appearing upon 
substitution of La with
Sr, the third corresponds to the positive shift due to 
the local increase of
the effective fractional charge on Cu. This expression 
agrees both
with the calculations in the frames of the 
ionic \cite{11a} as well as
of the cluster \cite{12} models  (in the uniform 
case $n=x$).

It follows from our results (Fig.3) that the 
resonance frequencies
$^{63}\nu_Q^{(1)}(x)$\ for line $\it{1}$ are 
shifted to lower values from the
reference value  $^{63}\nu_Q(0,0)=\,^{63}\nu_Q^0$ 
(we use here
$^{63}\nu_Q^0=31.9\ \rm MHz$\ estimated for 
$\rm La_2CuO_4$ \cite{10}). This
indicates that the positive   contribution to 
$^{63}\nu_Q(x,n)$\ is small and
that the effective fractional charge on sites 
$\it{1}$ is near zero.
It means that they are located in the regions 
free of doped holes. In contrast
line $\it{2}$ is due to the sites which in 
addition to the negative
shift exhibit a positive one.  It means that 
these sites belong to the regions
with an increased average charge (hole density) 
on the Cu ions.

The high frequency part of the spectrum can be 
analyzed by subtraction of the
$\it{1}$ and $\it{2}$ contributions from the 
entire signal.  The
resulting spectra are shown in Fig.1.  The 
frequencies corresponding to
their maxima are plotted in Fig.3.  We assume 
that this line
corresponds to the broadened 
$(\pm1/2) \leftrightarrow (\mp1/2)$
transitions of nuclei located in sites 
$\it{3}$ experiencing an
internal magnetic field (note the broad 
high-frequency shoulder).
The satellites are unresolved due to 
inhomogeneities of the
internal magnetic field and of the NQR frequencies.  
If the orientation of the
internal field with respect to the electric field 
gradient  is identical to that
observed for $\rm La_2CuO_4$ \cite{10}  
the frequency of this transition enables us  to 
estimate the quadrupole shift and to determine 
the Larmor frequency
for this Cu site to be 45.2 MHz for $x=0.12$. It 
corresponds to an internal 
field of 40.1 kOe.  Using the value of the hyperfine 
constant
$\left|A_Q\right|=139 \pm{10}\ \rm{kOe}/\mu_B$\ \cite{Imai} 
we estimate
that in order to create such a field the effective 
magnetic moment
of Cu at site $\it{3}$ has to be equal to 
0.29 $\pm{0.02}\,\mu_B$,
coinciding with the value obtained from neutron 
and muon
experiments\cite{5,13}.

Since quantitatively similar spectra were observed 
for each compound of the
first group we believe that they contain the same 
elementary 
"bricks" of the phase under study. Discussing the 
relative weight of
the different contributions, we note that 
the echo decay can be described in terms
of stretched exponents {exp}$[-(2t/T_2)^{a}]$ with 
different $T_2$ for
each site.  For $x=0.12$ the numerical fit of the 
measured echo decay
at the frequencies of different sites gives the 
same $a\simeq 0.5$
and $T_2^{(1)}=11\,\mu$sec;  $T_2^{(2)}=8.8\,\mu$sec;
$T_2^{(3)}= 5.5\,\mu$sec. Such a relaxation law is 
typical for the
relaxation via randomly distributed magnetic 
moments \cite{9}
whereas the values of  the relaxation rates depend 
on the location
of these moments with respect to different Cu sites.  
Extrapolating the
corresponding signal intensities to $t=0$ we find 
the contributions
of sites $\it{1}$, $\it{2}$, $\it{3}$ to be given 
by the ratio (1:6:13).

As for the origin of the sites $\it{1}$ it is possible 
to conclude that on one
hand they do not belong to the AF domains, and on 
the other hand they are outside 
of stripes since their effective charge is equal to 
zero. We assume that they
correspond to defects terminating the stripes. From 
their relative
number we estimate the average length of the stripe 
to equal at least 6 lattice constants.

It is important that the NQR frequencies for the 
site  $\it{2}$ (See Fig.4)
are almost the same for any $x$ thus indicating 
that for 
all Sr concentrations the
stripes are equally charged. The effective charge 
in a stripe is near
0.18-0.19.  This is larger than the average hole 
concentration (x)
but less than 0.5 expected for the ideal stripe 
picture \cite{1}.
It means that the charge is distributed over the 
domain wall
of a finite thickness.  Together with the above-mentioned 
intensity ratio
this indicates that the real stripe picture differs 
from the ideal one.
Another sign for this is the broadening of line $\it{2}$ 
due to a distribution of
NQR frequencies.  Its  linewidth (Fig. 3) reflects 
the behaviour of the
pinning: at $x=0.12$ where pinning is stronger, 
the narrowing due to the
motion of stripes is weaker and the linewidth is larger.  
The decrease of the
internal magnetic field with the deviation $x$ from
 0.12 reflects  the suppression
of magnetic order by holes penetrating into AF domains.

The changes in intensity of the NQR spectra are due to 
variation of the number
of "bricks" for the compounds with different Sr content, 
which 
depends on the pinning strength.
Our results indicate, that the stripe phase is pinned at 
least for the time
scales shorter than $10^{-6}$ sec (the same conclusion 
was reached also by La
NQR \cite{16a}).

The pinning for $0.08\leq x\leq 18$ is due to the 
buckling of the CuO$_2$
plane. It is connected with  the CuO$_6$ octaedra tilts 
around the [100] and [010]
axis by the angle $\Phi$,  which for given Eu 
substitution  is
governed by the Sr content. It follows from Fig. 2a, 
that the quantity of  the
pinned phase, which is proportional to the NQR signal 
intensity, is peaked at
$x=0.12$. This indicates additional strong pinning due 
to the commensurability
effect.  Such pinning is not unique for the LTT phase 
(as buckling is).
It is a manifestation of the plane character of the 
inhomogeneities of
the charge and spin distributions. Together with the 
existence of three
different Cu sites this gives an independent justification 
of the conventional
stripe picture \cite{1}, where the charges are uniformly 
distributed in rivers
of holes across one Cu-chain separated by the bare three 
leg ladders (we do
not discuss here the possibility of two magnetic sites 
which may be deduced
from the wide distribution of the internal field seen in Fig. 1).

Upon increasing $x$ over  $x=0.18$ the tilt angle is 
decreasing below the
critical value $\Phi_c\simeq 3.6^o$ \cite{10c} and 
depinning of the stripe
phase takes place.  Such behaviour occurs for the 
compounds with Sr
concentrations $x>0,18$ belonging to the second 
group for which the broad
signals, typical for the pinned phase, disappear.  
The corresponding NQR
spectra gradually transform to the narrow signal 
at higher
frequencies, which for $x=0.24$ is shown in the inset 
to Fig. 1.
The intensity of this line (proportional to the 
quantity of the
unpinned stripe phase) is shown in Fig. 2b.

The analysis of this relatively narrow signal reveals 
only two different sites
with $^{63}$Cu-NQR frequencies of 37.60 MHz and 
39.82 MHz. Within 1\% accuracy
these frequencies coincide with those known at the 
same $x$ for the $\it{A}$
and $\it{B}$ sites in the LTO superconducting 
phase \cite{11} confirming that
the LTT structure differs only in the directions 
of CuO$_6$ octaedra tilts.
The satellite $\it{B}$ is due to Cu having a localized 
hole in the nearest
surrounding since, according to \cite{12,12a}, its NQR 
frequency has the
additional positive shift $\delta\nu_Q\simeq 2.5\ \rm MHz$. 
The observed
transformation of the NQR spectra (in comparison 
with those for $0.08\leq
x\leq 0.18$) is due to the fast transverse motion 
of stripes in the depinned
phase. As a result the internal magnetic field on 
Cu nuclei is averaged out,
and the  effective fractional charge is homogeneously 
distributed over all Cu
nuclei giving the usual NQR frequencies.  Such depinning 
leads to the drastic 
changes in the magnetic properties. The echo 
signals decay 
for samples with $x>0.18$  becomes purely exponential 
($T_2^{(2)}=35.4\,\mu$sec for sample with $x=0.24$). 

An important feature of the compounds with 
$0.08\leq x\leq 0.18$ is the possibility to
observe the  NQR line in the state without bulk 
superconductivity. Usually for
$\rm La_{2-x}Sr_xCuO_4$\ compounds for moderate 
doping within the so-called
spin glass phase between $x\simeq 0.02$ and 
$x\simeq 0.06$ (the bulk
superconductivity threshold) the fast relaxation 
via the localized
moments\cite{9} hinders the observation of the 
Cu NQR.  In our
case the  Cu NQR  of  compounds moderately doped 
with Sr is observable even in
the absence of bulk superconductivity.  It is an 
indication that we are
dealing with the unusual correlated state where  
the magnetic moments created
upon Sr doping are not effective in relaxation. 
Note, that at 1.3K for $x=0.12$ 
compound the entire stripe phase is pinned. This 
follows from the   
comparison of the number of Cu nuclei, responsible 
for the NQR,  
with that for $x=0.24$ compound (Inset to Fig.1), 
which is due to 100\% of the Cu nuclei. Both 
these quantities 
were obtained by extrapolation the signals to 
$t=0$ and calculation
of the integrated intensities. 

It is also possible to make some remarks about the 
superconducting
properties. The main is that the depinning point 
separates two different types
of superconductivity.  For $x\leq 0.18$ we are 
dealing with a weak Meissner
effect, an increased London penetration length 
and with $T_c$ increasing
with  $x$ growing up to 0.18. Combining these 
facts with the
absence of a narrow signal typical for the bulk 
superconducting phase,
indicating that the impure  LTO phase is absent, 
and with the
suppression of the relaxation via magnetic moments 
of doped holes, one has
arguments in favor of possible one-dimensional 
superconductivity along the
charged rivers of stripes - the issue which is 
widely discussed
now \cite{14}.  For $x> 0.18$ we have bulk 
superconductivity with conventional
London length, typical NQR signal and decreasing 
$T_c(x)$.  Such crossover may
be caused by the transverse motion of the  
stripes carrying superconducting
currents which gives rise to the conventional 
superconductivity
in CuO$_2$ planes. Although possibly a simple 
coincidence, it happens when the
doping $x$ is equal to the effective charge 
$(n)$ in a stripe.

In conclusion we carried out Cu NQR studies of the Eu 
doped $\rm La_{2-x}Sr_xCuO_4$. We demonstrated that 
at 1.3K  
the ground state for moderate Sr content corresponds 
to the pinned 
stripe-phase and  that the pinning is enhanced at the 
commensurability. Three nonequivalent copper positions 
in the  
CuO$_2$   planes were found.  One of them with a  
magnetic moment  
of 0.29$\mu_B$  is related to the AF correlated 
antiphase domains. 
From the behaviour of the NQR frequencies  it 
follows that the 
effective charge of the domain walls separating 
these domains is 
almost independent on the Sr content $x$. The onset 
of the bulk 
superconducty at larger $x$ correlates with the 
dramatic transformation of the NQR spectra, indicating 
the 
depinning of the stripe phase.

The authors are grateful to H.Brom, A.Egorov and 
N.Garifyanov for valuable help and
discussions. This research was supported by the 
Deutsche
Forschungsgemeinschaft.  The work of G.T. was 
supported in part by the State
HTSC Program of the Russian Ministry of Sciences 
(Grant No. 98001)
and by the Russian Foundation for Basic Research 
(Grant No. 98-02-16582).



\begin{figure}
\caption{Representative Cu-NQR lineshapes at 1.3K of
$\rm La_{2-x-y}Eu_ySr_xCuO_4$\ with
$y=0.17$. The value of $x$ is shown for each line. All 
lineshapes include
standard frequency corrections of $1/\nu^2$ and are 
normalized
to equal heights. The raw data points are shown together
with their fits and the decomposition to different
contributions is also shown. The continuous line  
corresponds
to the contribution of sites $\it {1}$ and $\it {2}$. 
Filled circles show
the contribution of antiferromagnetic site $\it {3}$.  
Inset: A typical signal
for $x>0.18$ decomposed into two contributions (T=4.2K).
}
\end{figure}

\begin{figure}
\caption{The integrated intensity (normalized to maximal 
values)
of the Cu NQR signals for $\rm La_{2-x-y}Eu_ySr_xCuO_4$:
for $0.08\leq x\leq 0.18$ at 1.3K (a);
for $x\geq 0.18)$ at 4.2K (b).
}
\end{figure}

\begin{figure}
\caption{The different contributions to the Cu NQR signal for
the pinned stripe phase in $\rm La_{2-x-y}Eu_ySr_xCuO_4$\
as function of Sr content $x$ (T=1.3K):
$^{63}$Cu NQR frequencies of sites $\it {1}$ - open circles
and $\it {2}$ - filled circles; the corresponding half widths
at half maximum (HWHM) - open and filled triangles resp.;
the frequencies corresponding to the maxima of the magnetic
contribution $\it {3}$ are shown by squares.
}
\end{figure}

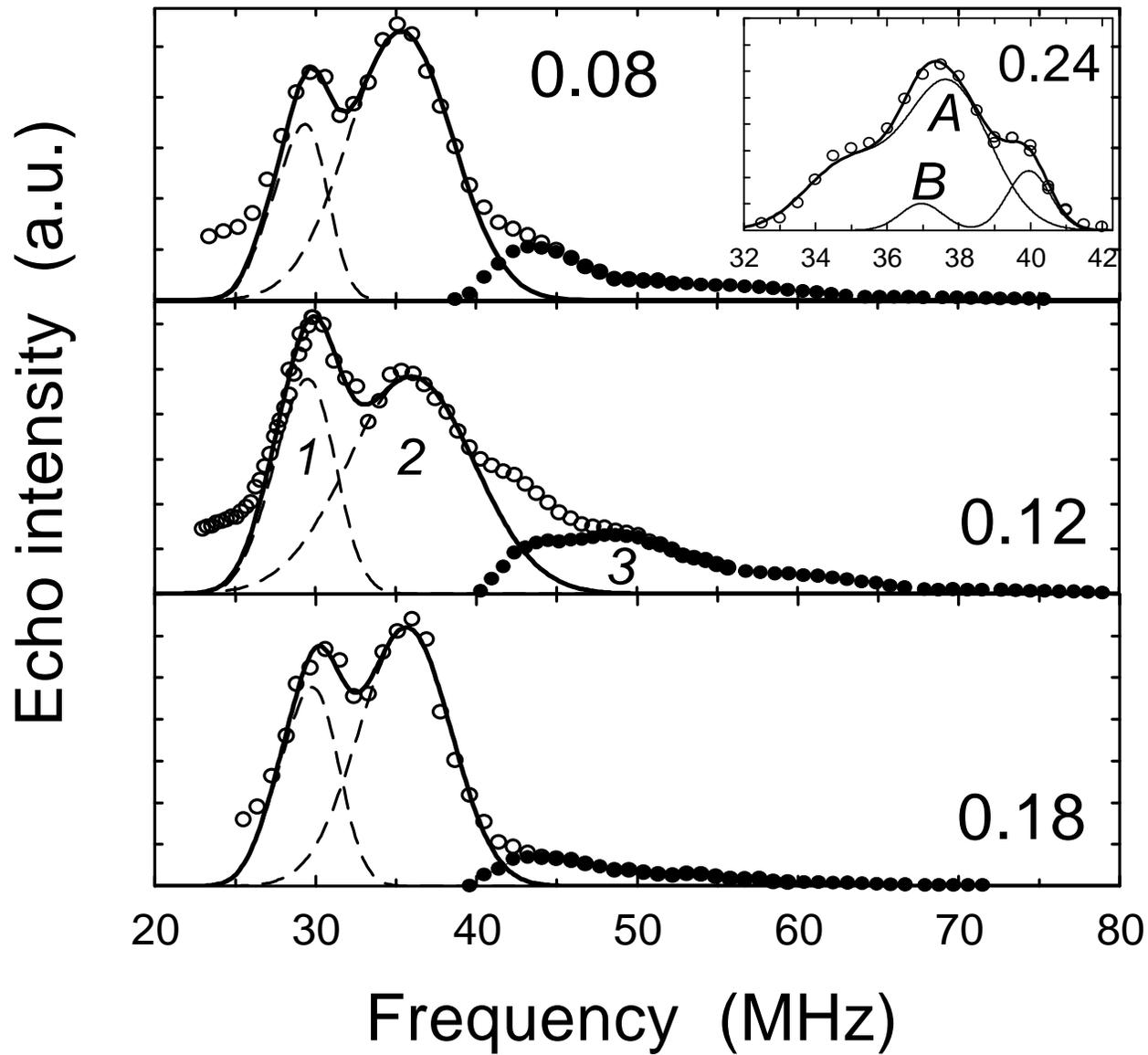
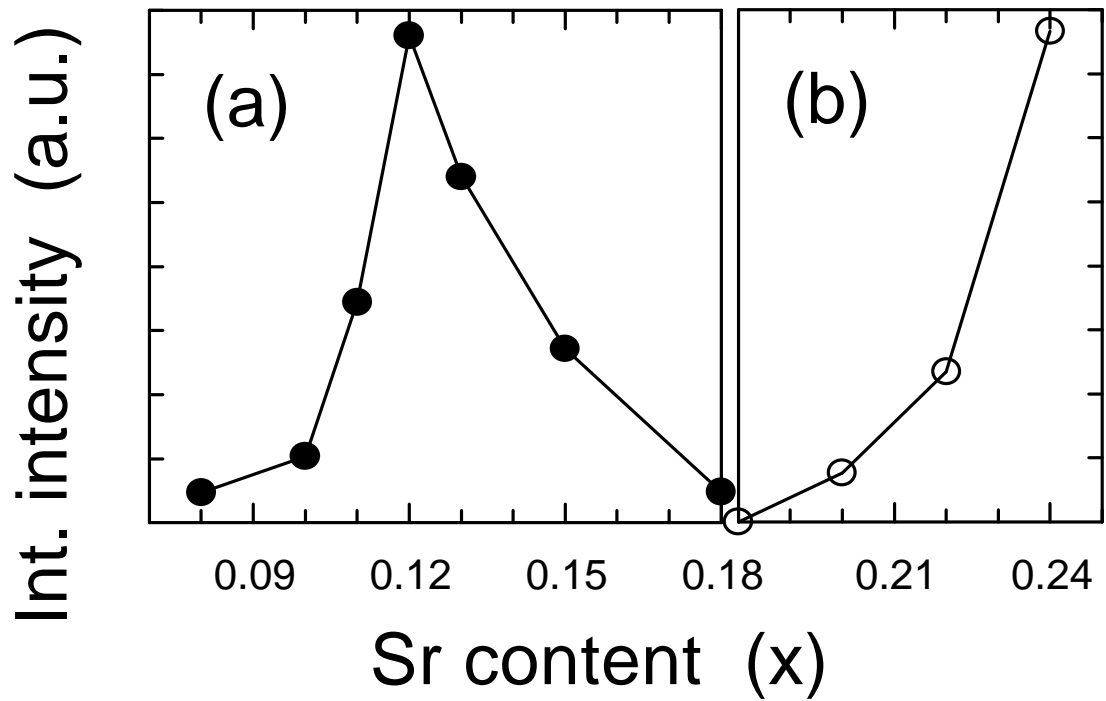
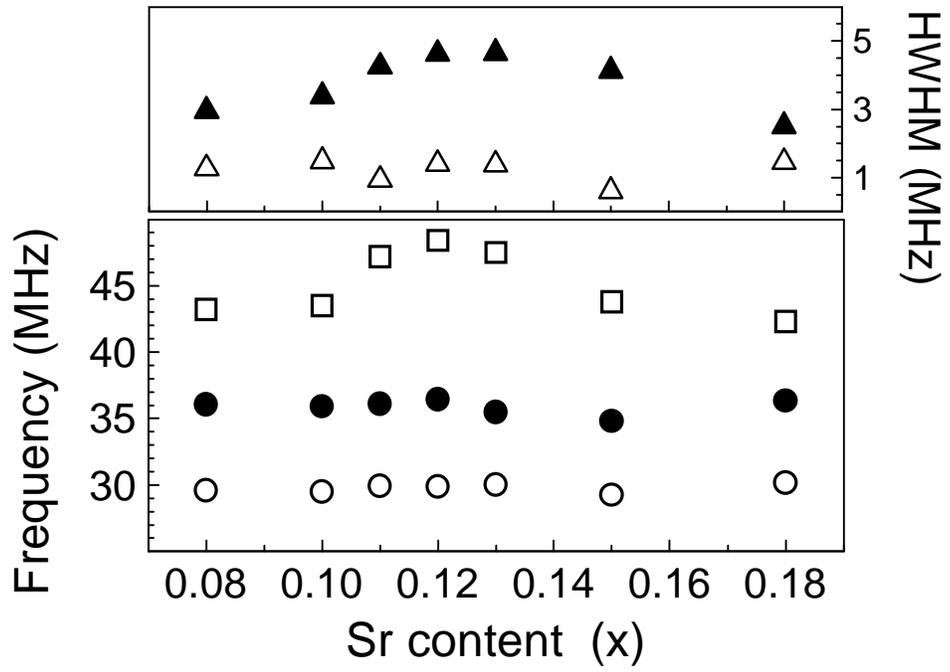
\end{document}